\documentstyle[11pt]{article}
\title{\bf Gott time machines in the Anti-de Sitter space}
\author{S\"oren Holst\thanks{Institute for Theoretical Physics,
    Stockholm University, Box 6730, S-113 85 Stockholm, Sweden.\newline
    e-mail: holst@vanosf.physto.se}}
\date{12 January 1995}
\begin{document}
\maketitle
\begin{abstract}
In 1991 Gott presented a solution of Einstein's field equations in 2+1
dimensions with
$\Lambda = 0$ that contained closed timelike curves (CTC's). This
solution was remarkable because at first it did not seem to be
unphysical in any other respect.
Later, however, it was shown that Gott's solution is tachyonic in a
certain sense. Here the case
$\Lambda < 0$ is discussed. We show that it is possible to construct
CTC's also in this case, in a way analogous to that used by Gott. We
also show that this construction still is tachyonic.

$\Lambda < 0$ means that we are dealing with Anti-de Sitter space, and since
the CTC-construction necessitates some understanding of its structure,
a few pages are devoted to this subject.
\end{abstract}

\section{Introduction}
\indent \indent General relativity in 2+1 dimensional
spacetimes is trivial in the sense that there
is no gravitation, that is masses in such spacetimes do not attract
each other [1]. This is due to the fact that in 2+1 dimensions
Einstein's equations (with $\Lambda = 0$) imply that spacetime is
flat in sourceless regions.
This, however, does not mean that 2+1 dimensional spacetimes are
uninteresting---in fact, despite the absence of gravitation the
subject is very rich, and yet it is
comparably easy to handle. One example of its richness was provided
1991 when Gott [2] showed how to construct closed timelike curves
(CTC's) by means of 2+1 dimensional masses. Since a
CTC allows an observer to travel backwards in time, and constitutes
what is generally meant by a time-machine, it was important to
investigate Gott's CTC-solution thoroughly (see $e.\: g.$ [3]). Soon it
was revealed that his construction was tachyonic---the system of 2+1
masses which give rise to the CTC's corresponds to an effective mass
travelling faster than light [4,5]. A stronger result, that every open
2+1 dimensional universe with $\Lambda = 0$ containing Gott-CTC's has
spacelike total momentum, also was shown [6,7]. The latter reference
generalizes the results of the former and also makes some statements
about the case $\Lambda < 0$. Furthermore, for closed
universes 't Hooft [8,9] showed that if one starts without CTC's and
tries to build them according to Gott's construction, then the
universe always will collapse before the CTC's come into existence.

In short, nature succeeds very well to prevent the creation of
time-machines. But these results concern the case with vanishing
cosmological constant $\Lambda$, and it is interesting to investigate
the situation when $\Lambda$ is non-vanishing. Is it perhaps not even
possible to construct a counterpart to Gott's CTC's in this case? This
and similar questions were the motivation for this work, in which we
repeat Gott's construction in the case $\Lambda < 0$ (that is, in the
Anti-de Sitter space) and then see whether the resulting CTC's spring
from a tachyonic system.

In the next section Gott's construction for the case $\Lambda = 0$
is reviewed, and we also show that this construction leads to a
tachyonic effective particle. Section 3 is devoted to Anti-de Sitter
space, and in section 4 we
perform the analogue of Gott's construction when $\Lambda < 0$. After
a lot of algebraic labour CTC's are found, and it is shown that the
condition for them to exist is essentially the same as in Gott's case.
In the subsequent section we go on to show that the CTC's only
exist for tachyonic systems of masses. This is done by a method
analogous to that used in the flat case.
The last section summarizes the results and contains a short
discussion.

As will be seen, the results in the case $\Lambda < 0$ parallell those
for $\Lambda = 0$ very closely. Since Gott's constrution is local,
and since locally there is no essetial difference between the cases, this is
perhaps only what one would suspect. But the CTC's generated also
exist infinitely far away from the masses involved and on a global
scale Anti-de Sitter space and Minkowski space differ very much. There
is for example no obvious reason for why the condition for CTC's to
exist in AdS-space is independent of their size.

\section{Closed timelike curves---Review of the situation with
  \mbox{$\Lambda = 0$}}
\indent \indent In 2+1 dimensions, as mentioned before, spacetime
is flat in sourceless regions, and the metric
exterior to a mass $M$ at the origin can be written [1]
\begin{equation}
  \label{Gottmet}
  ds^2 = -dt^2 + d\rho^2 + \rho^2 d\theta^2 \ ,
  \indent 0\leq\theta\leq 2\pi-\alpha
\end{equation}
where $\alpha = 8\pi GM$ ($G$ is the gravitational constant in $c = 1$
units). This means that a mass can be represented simply by cutting a
wedge out of ordinary Minkowski space, and identifying its sides. From
now on the angle of this wedge will be used to characterize masses.

We assume that the reader is already familiar with Gott's CTC
construction---here we only make a short description of it,
and give the condition for CTC's to arise. The construction simply
involves two masses $\alpha$ which are passing each other. If each of
them travels with a velocity greater than $\cos (\alpha/2)$ they will
give rise to CTC's. The intuitive reason for this is that the
identification of points at different sides of the cut out wedges,
will, for a mass in motion, involve a time jump. For $v > \cos (\alpha
/2)$ this jump in time will be large enough to permit a rocket
travelling around both of the masses to arrive to the starting point at
the same time that it left---that is, the time jumps will permit the
rocket to perform a CTC. See figure {\it 1}.
\begin{figure}
  \vspace*{5cm}
  \caption{To the left the Gott situation at $t=0$ is shown: Two
    equal masses $\alpha$ are passing each other. The regions inside
    each wedge are not part of the spacetime. The sides of the wedges
    are {\em not} identified in this plane since the masses are
    moving. The picture to the right shows the same situation in the
    full 2+1 dimensional spacetime. The dotted lines represent
    identified points on each wedge respectively.}
\end{figure}
There are different methods to show that the spacetime thus obtained
is tachyonic. To make the analogy with the Anti-de
Sitter space as clear as possible we will do it by using a
matrix formalism introduced by Deser, Jackiw and 't Hooft in 1984 [1], in
which any system of masses can be represented. The idea is to
represent a mass or a system of masses by the
Poincar\'e transformation that is the result of a parallell transport of
a vector around the masses.\footnote{In [1] the
Poincar\'e transformation representing a system of masses is regarded
as describing the ``matching conditions'' for points on different
sides of the cut out wedges. Then the formalism also contains
information about the relative position of the masses, but since this
information will be superfluous in this context we choose the
parallell transport viewpoint.}

To begin with, a mass $\alpha$ at rest is represented by an ordinary
rotation ${\bf R}(\alpha)$ of the Lorentz group in 2+1 dimensions,
since a vector parallell transported around the mass is
rotated by an angle $\alpha$. A moving mass $\alpha$ is represented by
${\bf L}(\xi){\bf R}(\alpha){\bf L}^{-1}(\xi)$ where ${\bf L}(\xi)$
with $\cosh \xi = \gamma_v = (1 - v^2)^{-1/2}$ is the boost of the
mass with respect to the laboratory. Namely, if we want to parallell
transport a vector around a moving mass, then we can always first make
the appropriate coordinate transformation to its restframe.

Now we want to represent the situation with the two masses in Gott's
construction in this way. In order to get as few matrices as
possible we choose to work in the frame of mass I (in figure 1). There
mass II is seen to pass by with velocity $u = 2v(1 + v^2)^{-1}$
and with matrices the situation is described as ${\bf R}_I{\bf
  L}_{II}{\bf R}_{II}{\bf L}^{-1}_{II}$, where ${\bf R}_I = {\bf
  R}_{II} = {\bf R}(\alpha)$ and ${\bf L}_{II} = {\bf L}(\xi)$ with
$\cosh \xi = \gamma_u$. Suppose now that we try to describe the
situation as only one effective mass instead of the two masses
$\alpha$, that is ${\bf L}(\xi_{eff},\theta){\bf R}(\alpha_{eff}){\bf
  L}^{-1}(\xi_{eff},\theta)$, where ${\bf L}(\xi_{eff},\theta)$
denotes a general boost in direction $\theta$. Since we want the
result of a parallell transport to be the same, independently of the
viewpoint taken, we get the equation
\begin{equation}
  \label{Ansatz}
  {\bf R}(\alpha){\bf L}(\xi){\bf R}(\alpha){\bf L}^{-1}(\xi) =
  {\bf L}(\xi_{eff},\theta){\bf R}(\alpha_{eff}){\bf
    L}^{-1}(\xi_{eff},\theta)
\end{equation}
If we use the trace of this equation together with its
time-time-component we can get the velocity (in the I-frame) of the
effective particle as an expression of $\alpha$ (each particle's mass)
and $u$ (the velocity of II in the I-frame). If we translate the
velocities to the laboratory (that is, the frame in which both masses
are seen to travel with velocity $v$ but in opposite directions) this
expression reads
\begin{equation}
  \label{Gottveff}
  v_{eff} = v \gamma_v \tan (\alpha /2)
\end{equation}

It is now easy to see that the Gott-condition $v > \cos (\alpha/2)$ is
equivalent to $v_{eff} > 1$. Hence we see that this type of CTC's
only exists when the system of two masses is tachyonic. As was mentioned
in the introduction stronger results than this have been shown, but
this is the one that will be generalized here to the case with
$\Lambda < 0$.

\section{The Anti-de Sitter space}
\indent \indent Just as Minkowski space is the vacuum solution to
Einstein's equations in 2+1 dimensions without cosmological constant
$\Lambda$, Anti-de Sitter space (or `AdS-space') is the
corresponding solution when $\Lambda < 0$. Here we discuss
the general properties of this spacetime to be able to redo the
Gott-construction. We will provide it with a special choice
of coordinates in which the geodesics are nicely represented, and when
discussing the isometries we show how to get explicit
expressions for them in these coordinates.

As definition of the AdS-space we will take the 3-surface of the
hyperboloid
\begin{equation}
  \label{hyperboloid}
  X^2 + Y^2 - Z^2 - T^2 = -1
\end{equation}
embedded in a 4-dimensional space with metric
\begin{equation}
  \label{4metric}
  ds^2 = dX^2 + dY^2 - dZ^2 - dT^2
\end{equation}
This surface can be parameterized in coordinates $r$, $\varphi$ och $t$
according to
\begin{equation}
  \label{XYZT(r,fi,t)}
  \left\{ \begin{array}{l}
    X = \sinh r \cos \varphi \\
    Y = \sinh r \sin \varphi \\
    Z = \cosh r \cos t \\
    T = \cosh r \sin t
  \end{array} \right.
\end{equation}
which yields the metric
\begin{equation}
  \label{trphimetrik}
  ds^2 = -\cosh^2 r dt^2 + dr^2 + \sinh^2 r d\varphi^2
\end{equation}

We see that $r$ and $\varphi$ plays the same role as ordinary polar
coordinates, while $t$ is the timecoordinate. But $t$ is periodic with
$2\pi$ and the topology is $S^1 \times R^2$. Such a spacetime, of
course, trivially contains CTC's and is therefore not what we are
interested in. Instead we will work with the universal
covering of the AdS-space, that is, the space where the time $S^1$ is
``unwinded'' to $R^1$ and $t$ is regarded as a normal non-periodic
variable. When we discuss the isometries, the space, strictly
speaking, is the ordinary AdS-space (with topology $S^1 \times R^2$),
but the result obtained is identical for its covering space.

If we transform the radial coordinate according to
\begin{equation}
  \label{rho(r)}
  \rho(r) = \frac{\cosh r - 1}{\sinh r}
\end{equation}
we obtain the metric
\begin{equation}
  \label{metric}
  ds^2 = - \left( \frac{1 + \rho^2}{1 - \rho^2} \right)^2 dt^2 +
  \frac{4}{(1 - \rho^2)^2} d\rho^2 + \frac{4\rho^2}{(1 - \rho^2)^2}
  d\varphi^2
\end{equation}
Observe that $\rho \rightarrow 1$ when $r \rightarrow \infty$, and in
the coordinates $(\rho, \varphi, t)$ the AdS-space is confined to a
cylinder, with the time lengthwise. The cross-section
of this cylinder is a $(\rho, \varphi)$-surface, and the point in
using these special coordinates is that a spacelike geodesic in such a
cross-section then always is an arc of a circle meeting the circle
$\rho=1$ at right angle. This means that a surface
$t$={\sc constant} looks like the the Poincar\'e model of a
space with constant negative curvature, and therefore we will call the
coordinates $(\rho, \varphi, t)$ the ``Poincar\'e coordinates''.
If we look at a $(\rho,t)$-plane of the spacetime cylinder we get
something like a Penrose-diagram, but with
non-straight lightlike geodesics---not with a constant slope of
$45^{\circ}$. (This is the price we have to pay for letting the
$t$={\sc constant} surfaces be Poincar\'e models.) Instead, lightlike
geodesics are described by the formula
\begin{equation}
  \label{lightlike}
  \rho(t) = \frac{1 - \cos(t+E)}{\sin(t+E)}
\end{equation}
where $E$ is a constant. Note that they traverse the whole space in
a finite time, namely $\pi$.

On the other hand, a timelike geodesic in a $(\rho,t)$-plane never
reaches infinity, but oscillates to-and-fro through the origin with a
period of $2\pi$. Two such geodesics that at time $t_0$ each is a
worldline of a resting object but at different places, a time $\pi/2$
later will focus at the origin, and now represent
the worldlines of two objects both at the origin, having different
velocities. This indicates a fundamental property of the isometries in
this spacetime, namely that an isometry which at time $t_0$ looks like
a translation, at time $t_0 + \pi/2$ has the effect of a boost.
We will come back to this in a moment. Geodesics in Anti-de Sitter
space: To the left in a $(\rho,t)$-plane and to the right in a
$(\rho,\varphi)$-plane.

Geodesics in ($\rho$,$t$)-planes and ($\rho$,$\varphi$)-planes are
illustrated in figure {\it 2}. A general timelike geodesic in the
AdS-space, spirals around the time axis $\rho=0$, but not necessarily,
of course, in a circular motion.
\begin{figure}
  \vspace*{5cm}
  \caption{Geodesics in Anti-de Sitter space: To the left in a
    $(\rho,t)$-plane and to the right in a $(\rho,\varphi)$-plane.}
\end{figure}

To be able to find and express the isometries it is now convenient to
parameterize the AdS-space by means of the group SU(1,1), in the sense
that every spacetime-point will correspond to a SU(1,1) matrix, and
conversely. Let a point $(X,Y,Z,T)$ on the hyperboloid (\ref{hyperboloid})
correspond to the matrix
\begin{equation}
  \label{coordmatris}
  {\bf K} \equiv \left( \begin{array}{cc}
               Z+iT & Y+iX \\
               Y-iX & Z-iT
             \end{array} \right)
\end{equation}
which belongs to SU(1,1) if $\det({\bf K}) = 1$.
Since $\det({\bf K}) = Z^2 + T^2 - Y^2 - X^2$ this is in
accordance with (\ref{hyperboloid}), and this relation between points
in the AdS-space and SU(1,1) matrices is 1-1.

The isometry group of 2+1 dimensional AdS-space is ${\rm SO(2,2)}$.
But this group is locally isomorphic to SU(1,1) $\times$ SU(1,1), and
we now claim that a general isometry can be written [6]
\begin{equation}
  \label{coordtrans}
  {\bf K' = T}_1{\bf K T}_2
\end{equation}
where {\bf K} is given by (\ref{coordmatris}) and ${\bf T}_1$, ${\bf
  T}_2$ are two arbitrary SU(1,1) matrices. To see this, write
\[ \left( \begin{array}{cc}
               dZ+idT & dY+idX \\
               dY-idX & dZ-idT
             \end{array} \right) = {\bf dK} \]
Now the metric can be written $ds^2 = -\det \, {\bf dK}$, and we see
that $\det({\bf dK'}) = \det({\bf T}_1{\bf dK T}_2) = \det({\bf dK})$
since ${\bf T}_1$ and ${\bf T}_2$ are SU(1,1). So the metric is preserved
and (\ref{coordtrans}) is indeed an isometry.

To begin with, let's find all isometries leaving the time axis
invariant, which means $t' = t$. Using (\ref{coordtrans}) together
with the coordinate transformations (\ref{XYZT(r,fi,t)}) and
(\ref{rho(r)}) gives that $t' = t$ if and only if the transformation
is of the form:
\begin{equation}
  \label{rotationalisometry}
  {\bf K'}={\bf RKR}^{-1} \hspace*{5mm} {\rm where}  \ {\bf R} =
  \pm \left( \begin{array}{cc} e^{i\theta/2} & 0 \\
                           0 & e^{-i\theta/2}
  \end{array} \right)
\end{equation}
The lower sign in {\bf R} only correspond to a mirroring of space,
and if we stick to the upper sign this transformation simply accounts
to a rotation around the time axis: $\rho' = \rho$, $\varphi' =
\varphi + \theta$ and $t' = t$.

Note that (\ref{rotationalisometry}) is the only isometry leaving the
time axis invariant. Hence, in AdS-space we do not have any ``pure''
translations, but of course it is possible to perform a
translation at one specified time.
Before we see how such an isometry is written in form
(\ref{coordtrans}) we will try to understand its properties. For this
purpose, look at figure {\it 2} and remember that an isometry maps
geodesics on geodesics. For example, by
performing a translation in the plane $t = 0$ we can map geodesic
$A$, which lies in $\rho = 0$, into geodesic $B$, which is at distance
$\rho_0$ from the origin at this time. But
in the $t = \pi/2$-plane $B$ passes $\rho = 0$ with velocity
$2\rho_0(1 + \rho_0^2)^{-1}$, as is not difficult to show, and hence
the performed isometry must have the effect of a boost there. In
fact, a translation of the planes $t = 2\pi n + \tau$
($n \in Z$) a distance $\rho_0$ automatically implies a boost of all
planes $t = 2\pi n + \tau + \pi/2$ with the mentioned velocity. We
will call isometries of this type for ``boost-translations'' and the
parameter $\rho_0$ for ``boost parameter''.

Lengthy calculations show that the boost-translation which is uniquely
specified by mapping $\rho = 0$ on $\rho' = \rho_0$, $\varphi' =
\theta$, in all planes $t = 2\pi n + \tau$ is written as
\begin{equation}
  \label{genboosttranslation}
  {\bf K'}=\frac{1}{1-\rho_0^2}\left( \!\! \begin{array}{cc} \vspace*{1mm}
             1 & i\rho_0 e^{i(\theta + \tau)} \\
             -i\rho_0 e^{-i(\theta + \tau)} & 1
           \end{array} \!\! \right) {\bf K}
           \left( \!\! \begin{array}{cc} \vspace*{1mm}
             1 & i\rho_0 e^{i(\theta - \tau)} \\
             -i\rho_0 e^{-i(\theta - \tau)} & 1
           \end{array} \!\! \right)
\end{equation}

For the sake of completeness we also write down the isometry
corresponding to a translation in time, that is $\rho' =
\rho$, $\varphi' = \varphi$ and $t' = t + \tau$:
\begin{equation}
  \label{timetransisometry}
  {\bf K'}={\bf TKT} \hspace*{5mm} {\rm where}  \ {\bf T} =
    \left( \begin{array}{cc} e^{i\tau/2} & 0 \\
                             0 & e^{-i\tau/2}
  \end{array} \right)
\end{equation}

By means of the rotations (\ref{rotationalisometry}), the boost-translations
(\ref{genboosttranslation}) and the time translations
(\ref{timetransisometry}) we can express an arbitrary isometry of
AdS-space, or in other words, we can map every two geodesics (both
either timelike or spacelike) into each other. In the following pages
we will need an explicit expression for a boost-translation in
the Poincar\'e coordinates. Since both rotations and time
translations are trivial in these, it will suffice to find such an
expression with $\theta$ and $\tau$ in
(\ref{genboosttranslation}) already fixed.
The following nasty expressions of the new coordinates
$(\rho',\varphi',t')$ in the old ones $(\rho,\varphi,t)$ and
the boost parameter $\rho_0$ gives the boost translation
(\ref{genboosttranslation}) with $\tau = \pi/2$ and $\theta = 0$.
 \begin{eqnarray}
  \label{rho'(rho,phi,t)}
  \rho' & \!\!=\!\! & \sqrt{\frac{A(\rho_0;\rho,\varphi,t) - (1 -
      \rho^2)}{A(\rho_0;\rho,\varphi,t) + (1 - \rho^2)}} \\
    & & {\rm with \ \ } A(\rho_0;\rho,\varphi,t) =
      {\scriptstyle \sqrt{(1+\rho^2)^2\cos^2t +
      \frac{1}{(1-\rho_0^2)^2}\left[ (1+\rho_0^2)(1+\rho^2)\sin t
      + 4 \rho_0\rho\cos\varphi \right]^2}} \nonumber \\
  \label{phi'(rho,phi,t)}
  \varphi' & \!\!=\!\! & \left\{ \begin{array}{ll}
    \arctan \left[ \frac{(1-\rho_0^2)\rho\sin\varphi}
    {(1+\rho_0^2)\rho\cos\varphi + \rho_0(1+\rho^2)\sin t} \right] &
    \ {\rm for} \ \cos\varphi > -\frac{\rho_0}{1+\rho_0^2}
    \frac{1+\rho^2}{\rho}\sin t \\
    \pi + \arctan \left[ \frac{(1-\rho_0^2)\rho\sin\varphi}
    {(1+\rho_0^2)\rho\cos\varphi + \rho_0(1+\rho^2)\sin t} \right] &
    \ {\rm for} \ \cos\varphi < -\frac{\rho_0}{1+\rho_0^2}
    \frac{1+\rho^2}{\rho}\sin t \end{array} \right. \\
  \label{t'(rho,phi,t)}
    t' & \!\!=\!\! & \left\{ \begin{array}{ll}
    \arctan \left[ \frac{(1+\rho_0^2)(1+\rho^2)\sin t +
      4\rho_0\rho\cos \varphi}{(1-\rho_0^2)(1+\rho^2)\cos t} \right] &
    \ {\rm for} \ -\frac{\pi}{2} < t < \frac{\pi}{2} \\
    \pi + \arctan \left[ \frac{(1+\rho_0^2)(1+\rho^2)\sin t +
      4\rho_0\rho\cos \varphi}{(1-\rho_0^2)(1+\rho^2)\cos t} \right] &
    \ {\rm for} \ \frac{\pi}{2} < t < \frac{3\pi}{2} \end{array}
  \right.
\end{eqnarray}

\section{Construction of CTC's}
\indent \indent Before we can try to construct any CTC's we must
understand what a mass looks like in 2+1 dimensional AdS-space.
This is carefully investigated in [10], but here it will suffice to use a
wedge representation corresponding to that used for $\lambda = 0$. As
in that case the spacetime exterior to sources is locally unaffected,
and a mass at rest at the origin can be represented in the Poincar\'e
coordinates simply as a wedge cut out according to figure {\it 3a},
with its sides identified at equal times. To find a more general
representation we just apply an isometry to this figure. For example,
the result of the isometry mapping point $A$ on point $B$ is shown in
figure {\it 3b}, and from this picture two things become apparent: 1)
At times when the mass is not situated at the origin the sides of the
wedges are not straight lines, but arcs of circles---the sides of the
wedges always follow spacelike geodesics. 2) The plane $t = 0$ in {\it
  3a} is after the transformation {\em not} a plane $t' = C$ for any
constant $C$, but the surface indicated in {\it 3b}. Since the
identification is done between points in this surface, it will involve
a time jump. This is in close analogy with the situation in figure
{\it 1}, where the identification also involves a jump in time. The
only difference is that in {\it 3b} the surfaces of identification
have different slopes at different times, and, for example, we get no
time jump at all at time $t' = -\pi/2$.
\begin{figure}
  \vspace*{5cm}
  \caption{a) To represent a mass $\alpha$ at rest at $\rho=0$ in
    AdS-space we simply cut a wedge out from the full spacetime. b)
    This shows the result from applying the boost-translation mapping
    point $A$ on $B$ in figure (a), and hence what a more general mass
    will look like.}
  \caption{A mass $\alpha$ at rest at $\rho=0$, and the space
    projection of the lightlike geodesic between $A$ and $B$ through the
    wedge. This also is the path connecting $A$ and $B$ in shortest time.}
\end{figure}

We will now investigate whether these time jumps can be made big enough
to give rise to CTC's, and in that case find a CTC-condition analogous
to $v > \cos (\alpha/2)$ in Gott's situation. The first step is to
find the path which gives the shortest time between $A$ and $B$ in
figure {\it 4} (The mass $\alpha$ is here at rest at the origin). This
path is a lightlike geodesic passing through the wedge, and its space
projection is shown in the picture. By applying
the boost-translation (\ref{rho'(rho,phi,t)}) -- (\ref{t'(rho,phi,t)})
on (\ref{lightlike}) (with $E = -\pi/2$), which is a lightlike
geodesic in the ($\rho$,$\varphi$)-plane, we get a more general
lightlike geodesic, passing the origin at distance $\rho_0 = \rho_c$.
Since the geodesic must meet the wedge in a straight angle, $\rho_c$ is
uniquely determined from $\rho_A$, and it is straightforward, but
algebraically tiresome, to find the time along this geodesic between
$A$ and $B$ expressed in $\alpha$ and $\rho_A$:
\begin{equation}
  \label{ABtime(rhoA,alpha)}
  t_{AB} = \pi - 2\arctan \left[ \frac{\sqrt{(1+\rho_A^2)^2 -
      4\rho_A^2\cos^2(\alpha/2)}}{2\rho_A\cos(\alpha/2)} \right]
\end{equation}

Now, put $t_A=-t_{AB}/2$ and $t_B=t_{AB}/2$. We want to find an
isometry mapping $t_A$ and $t_B$ on $t'_A$ and $t'_B$, such that
$t'_A = t'_B$. If such an isometry exists it is possible to travel
from $A$ to $B$ in no time (by applying the isometry to figure {\it 4}),
and we can repeat the argument in the
lower part of the figure with another mass subject to an opposite
isometry, which then will enable us to travel back again from $B$ to
$A$ in no time, and the CTC will be completed. Bearing Gott's case in
mind, the required isometry ought to be of the form
(\ref{rho'(rho,phi,t)}) -- (\ref{t'(rho,phi,t)}) since that will have
the effect of a boost in the $t=0$-plane and, as in figure {\it 3b},
therefore cause a time jump in the identification.

Using the time transformation, that is (\ref{t'(rho,phi,t)}), with
$\varphi_A = 0$, $\varphi_B = \pi$ we get the new time coordinates
$t_A'$ and $t_B'$ expressed in the old ones:
\begin{equation}
  \label{tA'(rho,phi,t)}
  t_A' = \arctan \left[ \frac{(1+\rho_0^2)(1+\rho_A^2)\sin t_A +
      4\rho_0\rho_A}{(1-\rho_0^2)(1+\rho_A^2)\cos t_A} \right]
\end{equation}
\begin{equation}
  \label{tB'(rho,phi,t)}
  t_B' = \arctan \left[ \frac{(1+\rho_0^2)(1+\rho_A^2)\sin t_B -
      4\rho_0\rho_A}{(1-\rho_0^2)(1+\rho_A^2)\cos t_B} \right]
\end{equation}
Demanding $t_B' < t_A'$ gives a condition on the boost parameter
$\rho_0$, and when inserting (\ref{ABtime(rhoA,alpha)}) into this one
finds that it is independent of $\rho_A$:
\begin{equation}
  \label{rho0>}
  \rho_0 > \frac{1-\sin(\alpha/2)}{\cos(\alpha/2)}
\end{equation}
If this condition is fulfilled we clearly can get CTC's, because if
$t_B' < t_A'$ we can, as already mentioned, repeat the argument with
another mass $\alpha$ in the lower halfplane of figure {\it 4}, for
the journey back from $B$ to $A$.

In (\ref{rho0>}) $\rho_0$ is the boost parameter of each mass, and as
mentioned before this means that their velocity as they pass each
other at the origin is $ \left. v \right|_{\rho=0} =
2\rho_0(1+\rho_0^2)^{-1}$. So (\ref{rho0>}) is equivalent to
\begin{equation}
  \label{CTC-condition}
  \left. v \right|_{\rho=0} = \frac{2\rho_0}{1+\rho_0^2} >
  \cos \frac{\alpha}{2}
\end{equation}
which is the same condition as in Gott's construction. However, in
AdS-space it is only true at $\rho = 0$ because here the geodesic
velocity is non-constant.

\section{The tachyonic nature of the CTC-spacetime}
\indent \indent As in the Minkowskian case we can associate an
isometry to every configuration of masses. Let a mass $\alpha$ at rest
(at $\rho = 0$) correspond to a rotation with angle $\alpha$, that is,
to an isometry (\ref{rotationalisometry}) with $\theta = \alpha$. In
analogy with ${\bf LRL}^{-1}$ for the flat case we then represent a
mass that is translated for example in the $t = 0$-plane, as ${\bf K' =
  BRB}^{-1}{\bf KB}^{-1}{\bf RB}$ where {\bf B} is the matrix in
(\ref{genboosttranslation}) with $\tau = 0$.
When $\Lambda = 0$ this representation of masses with isometries has
an immediate interpretation as the transformation resulting from a
parallell transport around the system. This interpretation is not
valid in the AdS-space, since in a curved space a parallell transport
along a closed loop {\em not} containing any masses yields a rotation
anyway. But actually, even though this interpretation does not work
when $\Lambda < 0$, the representation as such is still meaningful.

To get an equation analogous to (\ref{Ansatz}) we work in the frame of
mass I, where mass II is seen to pass by in direction $\varphi = \pi$ at
time $t = 0$. According to (\ref{genboosttranslation}) the
relative boost-translation between the masses then is described by
\begin{equation}
  \label{isomassI,II}
  {\bf K'}={\bf SKS}^{-1} \hspace*{5mm} ,
  {\bf S} = \frac{1}{\sqrt{1-\rho_1^2}} \left( \begin{array}{cc}
                                    1 & -\rho_1 \\
                                    -\rho_1 & 1
                                  \end{array} \right)
\end{equation}
where $\rho_1$ is the boost parameter of mass II in the I-frame. The
isometry corresponding to Gott's construction in AdS-space, seen from
mass I, then is
\begin{equation}
  \label{konstisometri}
  {\bf K'}={\bf RSRS}^{-1}{\bf K}{\bf SR}^{-1}{\bf S}^{-1}{\bf R}^{-1}
\end{equation}
The three pairs of matrices nearest to the {\bf K} account for mass
II, and the remaining rotation takes care of mass I.

Now we want to express the situation in terms of only one effective
particle. We do not know in which direction this effective
particle is travelling, so we have to use the boost-translation ${\bf
  S}_{\theta}$ that maps $\rho = 0$ on $\rho' = \rho_0$, $\varphi' =
\theta$ in every plane $t = 2\pi n + \pi/2$, that is
(\ref{genboosttranslation}) with $\tau = \pi/2$.
The isometry associated with one effective particle then is given by
\begin{equation}
  \label{effectivisometry}
  {\bf K'}={\bf S}_{eff}{\bf R}_{eff}{\bf S}_{eff}^{-1}{\bf K}
  {\bf S}_{eff}{\bf R}_{eff}^{-1}{\bf S}_{eff}^{-1}
\end{equation}
where ${\bf R}_{eff}$ is a rotation by some ``effective angle'' and
where ${\bf S}_{eff} = {\bf S}_{\theta}$ with boost parameter
$\rho_{eff,1}$. (The index 1 denotes that this effective boost
parameter is expressed in the I-frame.) Since (\ref{konstisometri}) and
(\ref{effectivisometry}) describe the same situation we get the
analogue of (\ref{Ansatz}) as
\begin{equation}
  \label{AnsatzAdS}
  {\bf RSRS}^{-1} = {\bf S}_{eff}{\bf R}_{eff}{\bf S}_{eff}^{-1}
\end{equation}
Some algebra yields, after translation of $\rho_1$ and $\rho_{eff,1}$
to the corresponding velocities in the laboratory (the frame in which
the two masses pass each other at $t = 0$ with equal velocities)
\begin{equation}
  \label{veff0(alpha,vL0)}
  \left. v_{eff} \right|_{\rho=0} = \frac{\left. v \right|_{\rho=0}
      \tan(\alpha/2)}{\sqrt{1-\left. v \right|_{\rho=0}^2}}
\end{equation}
This is the same expression as (\ref{Gottveff}), and since the
CTC-condition $\left. v \right|_{\rho=0} > \cos (\alpha/2)$ is the same too,
we have also in this case that $\left. v_{eff} \right|_{\rho=0} > 1$
exactly when the two masses generate CTC's: In AdS-space, as well as in
Minkowski space, we cannot have CTC's without permitting tachyonic
effective particles.

\section{Discussion}
\indent \indent We have shown explicitly that it is possible to
perform Gott's construction also in the Anti-de Sitter space, and that
it gives rise to CTC's when $\left. v \right|_{\rho=0} > \cos (\alpha/2)$.
This condition is the same as that shown by Gott for $\Lambda = 0$,
except for the fact that when $\Lambda < 0$ it concerns the velocity
of each particle only when they pass each other at the origin---the
velocities corresponding to Anti-de Sitter geodesics are non-constant,
but always largest at $\rho = 0$. We also showed that the
CTC-construction is tachyonic in the same sense as in the case
$\Lambda = 0$: A situation with this type of CTC's corresponds to an
effective particle with tachyonic momentum.

When $\Lambda = 0$ it is obvious that the CTC-condition must be
independent of how far away the CTC's are from the masses themselves,
since the construction with two wedges cut out from Minkowski space is
scale-invariant: Large CTC's look exactly as magnified small ones.
This is certainly not true in a curved spacetime, and particularly not
in AdS-space: The constant negative curvature has more effect on a
large scale than on a small one. Therefore it is a non-trivial fact
that the CTC-condition here also is independent on the largeness of
the generated CTC's.

The methods used here are highly specialized to AdS-space and hard to
generalize to other spacetimes. That the results are so similar to
those for $\Lambda = 0$ indicates that it should be possible to give a
more general treatment of the subject, and perhaps find a whole class
of spacetimes in which these or similar results hold.

\section*{Acknowledgements}
\indent \indent I would like to thank Ingemar Bengtsson for many
helpful discussions and for having drawn my attention to this subject.
I am also grateful to Karl Borg for some computational assistance
concerning the structure of AdS-space, and to Stanley Deser for a
comment.

\section*{References}
\newcounter{refnr}
\begin{list}
  {\arabic{refnr}}{\usecounter{refnr}}
  \item Deser,S., Jackiw,R., 't Hooft,G. (1984), ``Three-dimensional Einstein
    gravity: Dynamics of flat space'', {\it Ann.Phys.} {\bf 152}, 220.
  \item Gott,J.R. (1991), ``Closed timelike curves produced by pairs
    of moving cosmic strings: Exact solutions'', {\it Phys.Rev.Lett.}
    {\bf 66}, 1126.
  \item Cutler,C. (1992), ``Global structure of Gott's two-string
    spacetime'', {\it Phys.Rev. D} {\bf 45}, 487.
  \item Deser,S., Jackiw,R., 't Hooft,G. (1992), ``Physical cosmic strings do
    not generate closed timelike curves'', {\it Phys.Rev.Lett.} {\bf 68},
    267.
  \item Carroll,S.M., Fahri,E., Guth,A.H. (1992), ``An obstacle to
    building a time machine'', {\it Phys. Rev.Lett.} {\bf 68}, 263.
  \item Carroll,S.M., Fahri,E., Guth,A.H., Olum,K.D. (1994),
    ``Energy-momentum restrictions on the creation of Gott-time
    machine'',{\it Phys.Rev. D} {\bf 50}, 6190.
  \item Menotti,P., Seminara,D. (1994), ``Energy Theorem for 2+1
    dimensional gravity'', CTP \# 2324, gr-qc/9406016.
  \item 't Hooft,G. (1992), ``Causality in (2+1)-dimensional gravity'', {\it
    Class.Quantum Grav.} {\bf 9}, 1335.
  \item 't Hooft,G. (1993), ``Classical N-particle cosmology in 2+1
    dimensions'', {\it Class.Quantum Grav.} {\bf 10}, S79.
  \item Deser,S., Jackiw,R. (1984), ``Three-dimensional cosmological gravity:
    Dynamics of constant curvature'', {\it Ann.Phys.} {\bf 153}, 405.
\end{list}

\end{document}